\documentclass[onecolumn]{aastex631}

\usepackage{xcolor}
\usepackage{amsmath}
\usepackage{graphicx}
\newcommand{\eb}{\begin{equation}}
\newcommand{\ee}{\end{equation}}

\newcommand{\muas}{$\mu$as}

\definecolor{rkka}{RGB}{219,66,32}

\begin{document}

\title{Robust 1-norm periodograms for analysis of noisy non-Gaussian time series with irregular cadences:\\ Application to VLBI astrometry of quasars}

\author[0000-0003-2336-7887]{Valeri V. Makarov} \affiliation{United States Naval Observatory, 3450 Massachusetts Ave. NW, Washington, DC 20392-5420, USA}
\author[0000-0001-6759-5502]{S\'ebastien Lambert} \affiliation{SYRTE, Observatoire de Paris, Universit\'{e} PSL, CNRS, Sorbonne Universit\'{e}, LNE, 61 avenue de l’Observatoire 75014 Paris, France}
\author[0000-0002-8736-2463]{Phil Cigan} \affiliation{United States Naval Observatory, 3450 Massachusetts Ave. NW, Washington, DC 20392-5420, USA}
\author[0000-0001-5944-9118]{Christopher DiLullo} \affiliation{United States Naval Observatory, 3450 Massachusetts Ave. NW, Washington, DC 20392-5420, USA}
\author[0000-0001-8009-995X]{David Gordon} \affiliation{United States Naval Observatory, 3450 Massachusetts Ave. NW, Washington, DC 20392-5420, USA}

\begin{abstract}
Astronomical time series often have non-uniform sampling in time, or irregular cadences, with long gaps separating clusters of observations. Some of these data sets are also explicitly non-Gaussian with respect to the expected model fit, or the simple mean. The standard Lomb-Scargle periodogram is based on the least squares solution for a set of test periods and, therefore, is easily corrupted by a subset of statistical outliers or an intrinsically non-Gaussian population. It can produce completely misleading results for heavy-tailed distribution of residuals. We propose a robust 1-norm periodogram technique, which is based on the principles of robust statistical estimation. This technique can be implemented in weighted or unweighted options. The method is described in detail and compared with the classical least squares periodogram on a set of astrometric VLBI measurements of the ICRF quasar IERS B0642+449. It is uniformly applied to a collection of 259 ICRF3 quasars each with more than 200 epoch VLBI measurements, resulting in a list of 49 objects with quasi-periodic position changes above the $3\sigma$ level, which warrant further investigation.
\end{abstract}


\section{Introduction}
The Lomb-Scargle periodogram calculation is a powerful technique designed to reveal and characterize the periodic components in observational data sequences, which finds a wide scope of applications. For a review of its properties and underlying assumptions from the user's perspective, cf. \citet{2018ApJS..236...16V}. The need for this technique arises from the character of astronomical data (observational measurements), which are practically never evenly sampled in time. This makes the standard Fourier power spectrum analysis inapplicable for astronomical time series. Detection of orbiting exoplanets from precision radial velocities of host stars is one of the well-known use cases for the Least-Squares (LS) periodogram method \citep{2023AnRSA..10..623H}. The periodic component of the measured radial velocity sequence is caused by the reflex orbital motion of the host star orbiting the system's barycentre. The period of the main sinusoidal mode in the computed periodogram in this case estimates the orbital period of the planet, which often cannot be directly observed.

The periodogram method finds a somewhat less known application in precision astrometry of celestial bodies' positions. Binary stars with unresolved or dim companions have periodic signals, which are the harmonics of the orbital frequency, in either of the sky coordinates referenced to a fixed celestial frame. Given a significantly long and precise cadence of position measurements covering at least one orbital period, the more general approach is to directly fit a set of Kepler elements of the emerging explicitly nonlinear 2D model, which proves a daunting and ambiguous task in the presence of even a small admixture of statistical outliers \citep{2006ApJS..166..341G}. A robust and reliable periodogram decomposition is a welcome alternative when a large amount of observational data has to be processed with a low output of true positives. The need for a resilient periodogram algorithm, which can produce meaningful results outside of the normal distribution of data points, also emerges in the interpretation of high quality photometric time series. Magnetically active stars, for example, often manifest complex structures of signals in their light curves with periodic modulation mixed with stochastic, unpredictable bursts of radiation \citep{2017ApJ...845..149M}. 

Our main goal for this study is to develop and test a modification of the classical 2-norm periodogram algorithm (also known as the Lomb-Scargle periodogram) based on the principles of robust statistical estimation. This algorithm is intended to be used for processing of a massive data base that includes single-epoch astrometric measurements of thousands of radio-emitting quasars with the geodetic Very Long Baseline Interferometry (VLBI) world-wide facility. The system of accurate positions of these sources constitutes the fundamental International Celestial Reference Frame \citep[ICRF3, ][]{2020A&A...644A.159C}, which underpins all other derivative celestial and geodetic reference frames. The astrometric stability of the most frequently observed quasars is of crucial importance for the overall accuracy and stability of ICRF3. We therefore develop a method to determine if some of the ICRF3 sources manifest periodic signals in their celestial positions, which could emerge from dual orbiting black holes in their centers, as well as a number of other
effects in the extended structures and jets  \citep{2012MmSAI..83..952M}.

The need for robust statistical estimation techniques generally arises in astronomical data processing when the available data are ridden with a large fraction of outliers outside of the commonly assumed Gaussian distribution of errors, representing a heavy-tailed sample distribution. Examples of critically important applications can be found in the mutual orientation alignment of different celestial reference frames  \citep{2021MNRAS.506.5540M, 2023A&A...669A.138L, 2023AJ....165..202F}, where common object show a high rate of position offsets with extremely low formal probabilities.

\section{The least-squares (2-norm) periodogram}
\label{ls.sec}

In the most general setup of the problem, our task is to mathematically analyze a given time series (observations) $d(t_i)$, where periodic sinusoidal signals may be hidden. The data is discretized on a sequence of specific times of measurement $t_i$, $i=1,2,\dots ,N$. When the measurements are taken on a regular, equally spaced grid with time step $\Delta t$, the problem is solved by the direct Fourier transform and subsequent computation of the Fourier power spectrum. The spectrum is quantified on a grid of angular frequencies
$f_k=2\,\pi /(k\Delta t)$, $k=2,3,\dots,N$, where the highest non-degenerate frequency $f_2=\pi/\Delta t$ is the angular Nyquist frequency. This set of frequencies is complete, because all other signals within the functional space spanned by the Fourier basis functions are not independent. In other words, the fitting function
\eb
\hat d(t_i)\equiv \hat d_i=\sum_{k=2}^{N} \left(c_k\,\cos(f_k\,t_i)+s_k\,\sin(f_k\,t_i)\right)
\label{model.eq}
\ee
is the exact and unique representation of any sequence $d_i$ with a zero mean. This is no longer true if the cadence $\{t_i\}$ is irregular. The Fourier harmonics are not orthogonal if sampled on an irregular cadence. In principle, this difficulty can be bypassed by constructing an ad hoc orthogonal basis from the Fourier harmonics using, for example, the Gram-Schmidt process, but the practical value of such representation is dubious, because the emerging fitting functions do not find a simple interpretation. However, we can disregard the issue of nonorthogonality and seek a solution to Eq. \ref{model.eq} for a chosen $f_k$. Under this generalization, the classical periodogram analysis is equivalent to the least-squares (LS) fitting of model (\ref{model.eq}) for a grid of trial periods $p_k=2\,\pi/f_k$ \citep{1982ApJ...263..835S}. The emerging LS problems, for a specific $p_k$, can be written as
\eb 
\boldsymbol A\cdot \boldsymbol x = \boldsymbol d,
\label{a.eq}
\ee
where the design matrix $\boldsymbol{A}$ has two columns with calculated sequences of $\cos(f_k\,t_i)$ and $\sin(f_k\,t_i)$ values, and the number of its rows is equal to the number of data points $N$. The right-hand part vector $\boldsymbol{d}$ is the vector of centralized observations, and the vector of unknowns $\boldsymbol{x}$ comprises the two coefficients $c_k$ and $s_k$. Standard LS algorithms can quickly solve this system  to obtain the solution vector
\eb 
\label{ls.eq}
\hat{\boldsymbol{x}}=(\boldsymbol{A}^T \boldsymbol{A})^{-1}\,\boldsymbol{A}^T\cdot \boldsymbol{d}.
\ee 
Any standard LS algorithms can be used to solve this system. This has to be done for each trial period $p_k\equiv 2\pi/f_k$ separately, including the setup of the design matrix $\boldsymbol{A}$. In the example used in our paper, the number of trial periods $N_p=1000$, but in other applications, it can be up to $O(10^5)$. This is not a problem for modern computers and LS algorithms, but in the 1970-ies, when the periodogram method started to attract astronomers' attention, the speed of computation was a crucial consideration. This motivated \citet{1976Ap&SS..39..447L} to propose a modification where the design matrix $\boldsymbol{A}$ is orthogonalized by introducing a phase shift $\tau_k$ in each of the fitting functions in (\ref{model.eq}), to the effect that the normal matrix $\boldsymbol{A}^T\,\boldsymbol{A}$ becomes diagonal. This modification has little practical advantage now but it brings in additional restrictions precluding necessary extensions of the model, as we will now discuss. Lomb's modification is therefore not recommended. A more accommodating and rigorous way of orthogonalization, if such a technical action is deemed desirable, was proposed by 
\citet{1981AJ.....86..619F}. \citet{1982ApJ...263..835S} also mentions that the spectral power of a pure noise data has a more predictable statistical distribution, but this argument is only valid for Gaussian noise with equal variances, which is never the case in practice.

\section{Extended periodogram models}
Astronomical data series $\boldsymbol{d}$ often include non-periodic components. Using the archetypical example of detection of exoplanet signals in precision radial velocity measurements of host stars, the expected additional terms include a constant offset (the systemic radial velocity of the exoplanet system) and possibly a linear trend from perspective acceleration or a distant binary companion. It is not recommended to estimate these terms separately and prior to the periodogram solution and subtract them from the original data, which, unfortunately, is often done in practice. The reason why this pre-processing leads to an error in the periodogram is that these terms are not orthogonal to the fitted sine functions on a non-uniform cadence of data points. Unlike the regular Fourier transform, the trial frequencies are not integer multiples (harmonics) of the time interval. If a sinusoidal signal is present in the data, its estimated amplitude or power will be affected by the biased estimate of the constant term. The only correct and consistent way of dealing with additional terms is to include them in the fitting model \citep{1981AJ.....86..619F, 1999ApJ...526..890C}. For example, the fitting model suitable for exoplanet detection can be
\eb
\hat d_i=x_0+x_1(t_i-t_0)+x_2\,\cos(2 \pi\,t_i/p_k)+x_3\,\sin(2 \pi\,t_i/p_k)
\label{model2.eq}
\ee
for each trial period $p_k$. Note that a separate constant term $x_0$ and a linear slope $x_1$ are obtained for each trial period, and they are not equal for different trial periods. The variation of these terms with the trial period reflects the error introduced into the periodogram by subtracting the common terms a priori.

The periodogram estimate can be the amplitude of the fitted sinusoid
\eb 
a(p_k)=(x_2^2+x_3^2)^\frac{1}{2},
\label{amp.eq}
\ee 
or the power
\eb 
s(p_k)=x_2^2+x_3^2.
\ee 
As discussed above, we are using the amplitude periodogram in this paper, which has a more intuitive interpretation as the amplitude of the periodic signal in the same units as the measurements. The linear condition equations still take the form (\ref{a.eq}), but the design matrix $\boldsymbol{A}$ now has four columns, and vector $\boldsymbol{x}$ includes four unknowns $x_0$, $x_1$, $x_2$, and $x_3$. The solution vector $\hat{\boldsymbol{x}}$ is obtained from the least-squares solution, Eq. \ref{ls.eq}. If a significant signal $a(p_k)$ is detected, the corresponding values $x_0(p_k)$, $x_1(p_k)$ provide the best estimates of the constant term and the linear trend. 

\section{Statistical uncertainties}
\label{unc.sec}
What is the confidence level of a detected signal in the LS periodogram? This is an estimate of crucial importance, because the probability of the null hypothesis (that the detected feature is just a random fluke), also known as the false alarm probability (FAP), determines if we can believe the result. Traditionally, a high formal confidence is desired in astronomical applications such as detection of exoplanet signals, the recommended value being 0.997 (the $3\sigma$ level for a normal distribution). A robust method of estimating the FAP is the 
bootstrap simulation, which is also extendable to non-Gaussian distributions of measurement error. This is the method of ``last resort'' when the signal-to-noise ratio (SNR) of the detected signal leaves room for a catastrophic false positive. It is computationally expensive, however, and requires a sufficiently large number of data points. Monte Carlo methods, which are also computationally expensive, can be efficient when the statistical distribution of the observational noise is known. A random number generator is used to construct a sequence of synthetic measurements on the given sequence of times $t_i$, then a periodogram solution is obtained for each realization of noise, and the signal amplitude $(c_k^2+s_k^2)^\frac{1}{2}$ is computed. Repeating this process multiple times ($O(10^3)$ is usually required for an accurate estimation) allows us to estimate the CDF of the posterior distribution of the periodogram amplitude at any trial period, and hence, the $p$-value of the null hypothesis (or FAP). 

Here we describe a computationally efficient and direct method of confidence estimation for LS periodograms in the extended form (Eq.~\ref{model2.eq}). Noting that the standard periodogram solution obtained from the LS adjustment
per Eq.~\ref{ls.eq} is already based on the assumption that the measurement error is normally distributed, a direct computation of the periodogram cumulative distribution function (CDF) can be performed. If the covariance matrix $\boldsymbol{C}_d \equiv E[\boldsymbol{d}\cdot \boldsymbol{d}^T]$ is known or assumed, the corresponding covariance of the solution vector is
\eb 
\label{cx.eq}
\boldsymbol{C}_x \equiv E[\hat{\boldsymbol{x}}\,\hat{\boldsymbol{x}}^T]=
\boldsymbol{C}\,\boldsymbol{A}^T\,\boldsymbol{C}_d\, \boldsymbol{A}\,\boldsymbol{C},
\ee 
where $\boldsymbol{C}=(\boldsymbol{A}^T\boldsymbol{A})^{-1}$. It is often assumed that the measurements $\boldsymbol{d}$ are statistically independent, in which case the matrix $\boldsymbol{C}_d$ is diagonal. If the errors also have the same variance $\sigma^2$, this equation further simplifies to $\boldsymbol{C}_x=\sigma^2\,\boldsymbol{C}$.

The solution covariance $\boldsymbol{C}_x$ is a $4\times 4$ symmetric matrix, which is computed for each trial period $p_k$. We are mostly interested in the $a(p_k)$ statistics per Eq. \ref{amp.eq}. The two involved statistics $x_2$ and $x_3$, in accordance with the assumed normal distribution ${\cal N}(0,\sigma_i)$ for each data point, are binormal variates, whose covariance matrix $\boldsymbol{C}_a$ is the corresponding $2\times 2$ block of $\boldsymbol{C}_x$. The estimated vector $\boldsymbol{y}=[x_2,x_3]^T$ can be standardized to obtain 
\eb 
\bar{\boldsymbol{y}}=\boldsymbol{C}_a^{-\frac{1}{2}}\,\boldsymbol{y},
\ee 
so that $\bar{\boldsymbol{y}}$ is a binormal uncorrelated variate of unit variance. This is equivalent to determining the error ellipse for binormal variates. The components of $\bar{\boldsymbol{y}}$ can be interpreted as the upper and lower signal-to-noise ratios of the given periodogram result. Consequently, the quadratic form
\eb 
\psi=\bar{\boldsymbol{y}}^T\bar{\boldsymbol{y}}=\boldsymbol{y}^T\boldsymbol{C}_a^{-1}\,\boldsymbol{y}
\label{psi.eq}
\ee 
is a $\chi^2$-distributed variate with 2 degrees of freedom. The corresponding confidence of rejecting the null hypothesis can be computed from the cumulative distribution function (CDF) of the distribution $\chi^2[2]$ for each periodogram value. For a graphical representation of perodogram results, it is convenient to compare the confidence levels to specific points, which correspond to the $\pm 1\sigma$, $\pm 2\sigma$, and $\pm 3\sigma$ intervals of the normal distribution, which have the cumulative probabilities of 0.683, 0.955, and 0.997, respectively. The corresponding levels of $\psi$ (computed as CDF$^{-1}\left[\chi^2[2]\right]$) are
2.296, 6.180, and 11.829. Periodogram amplitudes with $\psi$-values above 11.829 can then be regarded as highly confident positive detections at a confidence level above 0.997.

In the exoplanet detection literature, an alternative method of FAP-estimation is often used, developed by \citet{2008MNRAS.385.1279B}. It is also based on the assumption that the signal contains only a finite set of model (base) functions, and the random component of the data vector is pure Gaussian noise with the known standard deviations $\sigma_i$. The statistical significance of a single periodogram value can then be naturally estimated from the properly normalized difference of the reduced $\chi^2$ statistic of residuals with and without the corresponding harmonic terms \citep[e.g., Eq. 6 in][]{1999ApJ...526..890C}, which follows the $F$-distribution (or beta-distribution if two or more specific periodogram frequencies are considered). However, while we have in practice a large number of periodogram value realizations, only the maximum value and the corresponding trial period are of interest. Even in the absence of detectable signal, given a large number of trials, it is probable that the highest significance value exceeds the threshold confidence level. This probability can be estimated within the extreme value statistic of an $F$-distributed homoscedastic random process assuming that the periodogram values are independent. This method is vulnerable to aliasing, which is caused by the limited spectral window of the given data series. The adjacent periodogram values are not independent, and periodogram features become increasingly wider ``window functions" toward the longest trial periods. Non-uniform cadences with long gaps can also generate aliasing, spectral leakage, and spurious periodogram peaks. The extreme-value distribution fitting method \citep{2014MNRAS.440.2099S} is more general for non-Gaussian processes, but it still refers to the null hypothesis of uncorrelated white noise in the data, which is inaccurate for the specific applications in this paper, or the spectroscopic detection of exoplanets \citep{2009ApJ...707L..73M}.

\section{To weight or not to weight?}

Astronomical time series often have unequal formal errors of individual data points. The formal error represents the expected standard deviation of the measurement, which can vary in a wide range because of observational conditions, instrument setup, etc. The LS solution in Eq. \ref{ls.eq}, on the other hand, is unweighted, because it does not involve the estimated formal errors. The standard way of dealing with processing data of non-uniform precision is to use weighted LS fitting. It can be applied to LS periodogram analysis too in the framework of weighted periodogram solution. The basic equation replacing (\ref{a.eq}) becomes
\eb 
\boldsymbol W\,\boldsymbol A\cdot \boldsymbol x = \boldsymbol W\,\boldsymbol d,
\ee 
where the weight matrix $\boldsymbol W= \boldsymbol{C}_d^{-\frac{1}{2}}$. The covariance of the right-hand part is now the identity. The formal covariance of the periodogram coefficients of interest transforms from Eq. \ref{cx.eq} into
\eb 
\label{cw.eq}
\boldsymbol{C}_x =
\left(\boldsymbol{A}^T\,\boldsymbol{W}^2\, \boldsymbol{A}\right)^{-1}.
\ee 
The subsequent analysis of periodogram uncertainties is the same as described in Sect. \ref{unc.sec}. 

We performed limited experiments using the formal weights on the example described in Section \ref{why.sec} to estimate the impact of this additional modification. We found rather limited changes in the computed periodogram amplitudes with the unweighted and weighted LS options. The most prominent features indicating possible signals have approximately the same shape and location. The greatest difference is found in the estimation of the $1\sigma$ and $3\sigma$ confidence intervals. The weighted covariance of the periodogram coefficients $\boldsymbol{C}_a$ is generally much smaller for the weighted solution than for the unweighted solution. This is caused by a large spread of individual formal errors, and the fact that the weighted LS solution is optimal. If all the formal errors are equal, the periodogram covariances and the derived amplitudes become equivalent in the two solutions. Thus, the weighted covariance $\boldsymbol{C}_a$ is the global minimum of all possible unweighted counterparts. The lower covariances result in narrower confidence intervals, and the net result is that most of the periodogram solution becomes a highly confident positive detection. This result is completely misleading for the given example, because, as we will see in the next Section, the formal errors of the data points have little bearing on the actual dispersion and statistical distribution of the data.

\section{Why do we need something else?}
\label{why.sec}
Let us summarize the implicit assumptions involved in the LS periodogram method. 
\begin{enumerate}
    \item 
    The data vector is a composition of random uncorrelated noise and a single monochromatic sinusoidal signal, whose amplitude and period are to be determined.
    \item 
    The measurement noise is Gaussian.
    \item 
    The data sequence is centralized, i.e., has a zero mean---unless the extended version of the method is employed.
\end{enumerate}

An example when the second assumption is violated can be found in \citep{2010ApJ...717.1202M}. The astrometric position (photocenter) of the Sun as measured by a distant observer is subject to stochastic variations caused by the presence of sunspot groups and bright plage areas on the rotating surface. Each photometric feature generates a time-variable shift of the unresolved disk on the time-scale of days, which is not periodic because of the phase scrambling. The composition of such stochastic signals is an unpredictable ``jitter''. The measured shifts from the mean photocenter show an utterly non-Gaussian distribution because the intrinsic distributions of the sunspot sizes, lifetimes, and positions within the disk are not normal. In this case, the nominal LS periodogram, as well as the traditional FAP estimation, are likely to produce misleading and inaccurate results.

\begin{figure*}
\centering
\includegraphics[width=.45\textwidth]{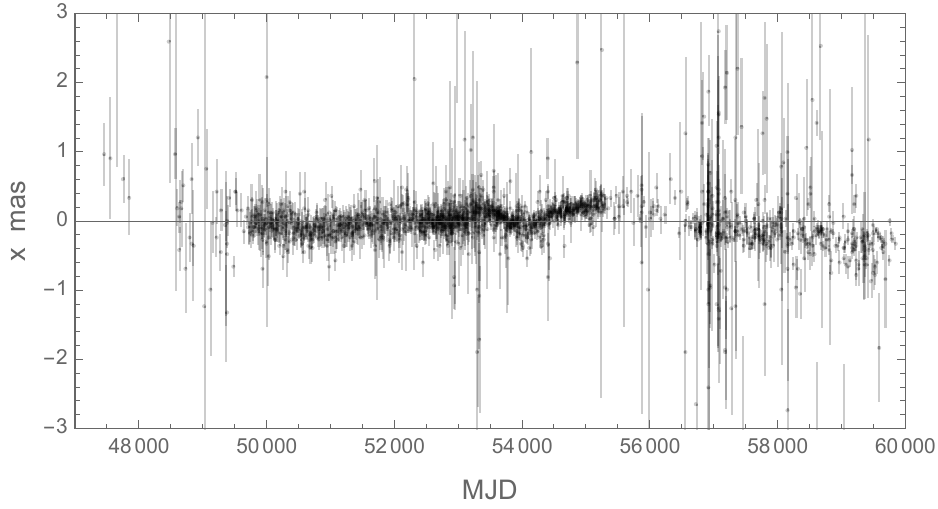}
\includegraphics[width=.45\textwidth]{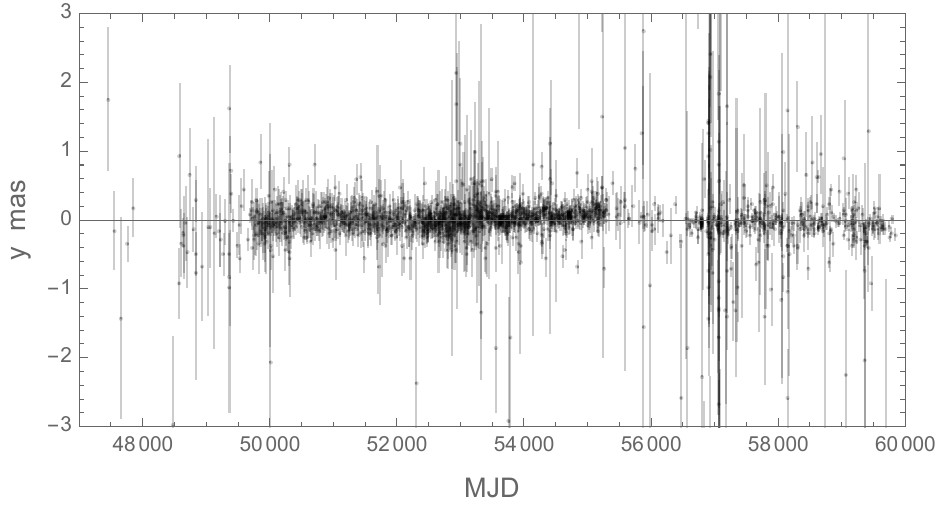}
\caption{Astrometric offsets from the mean position of the ICRF3 source IERS B0642+449 measured by VLBI over 30 years. Left plot: right ascension tangential component ($x$) in mas.
Right plot: declination tangential components ($y$) in mas. Each data point is shown with its formal $\pm 1\sigma$ error bar. \label{data.fig}}
\end{figure*}

Fig. \ref{data.fig} shows the observed time series used in this paper to illustrate the application of the proposed 1-norm periodogram analysis. It shows the high-accuracy astrometric data collected by geodetic VLBI for the ICRF3 source IERS B0642+449 for nearly 40 years of continuous observations. Each data point corresponds to a one-day ``session'' with multiple delay measurements of this source together with a number of other ICRF3 sources. The observational data are represented as coordinate offsets $x=(\alpha_{\rm obs}-\alpha_{\rm mean})\,\cos \delta_{\rm mean}$ (left panel) and $y=\delta_{\rm obs}-\delta_{\rm mean}$ (right panel) in mas, where
$\{\alpha_{\rm mean}, \delta_{\rm mean}\}$ are the weighted mean coordinates for this source in the equatorial coordinate system. The formal errors for each observation are shown as error bars. This enigmatic high-redshift ($z=3.41$) gamma-ray blazar is obviously one of the astrometrically unstable ICRF3 sources with shifting position mostly in the R.A. component. The origin of the position variations is outside of the topic of this paper, but we briefly note the study by \citet{2016AJ....152..151X}, who detected a dual structure of IERS B0642+449 from a closed delay analysis of a high-intensity geodetic VLBI session. The detected separation of the dual components is approximately 0.46 mas and the position angle is $262.2\degr$. One of the interesting applications of sub-mas astrometry with VLBI is the possibility of detection of orbiting dual AGNs. With an estimated scale of 7.6 pc/mas at this redshift, a binary black hole with a total mass of $10^{10}M_{\sun}$ and a period of 40 years may have an angular separation of about 10 \muas, which may be within the reach with this type of data. Central engine binarity may be one of the explanations for the observed quasi-periodic modulation of some gamma-ray blazars' light curves \citep{2015ApJ...813L..41A}.

Both coordinate trajectories in Fig. \ref{data.fig} appear to include long-term variations and, possibly, periodic components on the timescale of a few hundred days. Are they statistically significant? We begin with the standard LS periodogram analysis using the extended model Eq. \ref{model2.eq}. The need to include the linear terms, in particular, comes from the possibility of a ``secular'' proper motion in the data, which is not part of the astrometric model used in the VLBI data reduction pipeline. We compute the periodogram fitting coefficients $\{x_0,\ldots,x_3\}$ for an exponential grid of 1000 trial periods, $p_k=p_0\,{\rm dex}(q\,k)$, with the exponent step $q=(3652.5-p_0)/1000$ in days. The longest trial period is then 10 yr, which is practically limited by the time span of the available data. 

\begin{figure*}
\centering
\includegraphics[width=.45\textwidth]{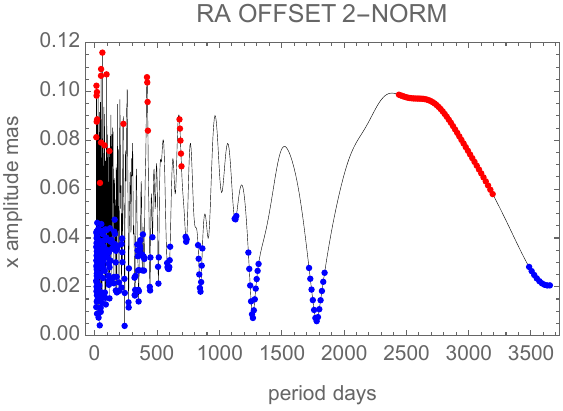}
\includegraphics[width=.45\textwidth]{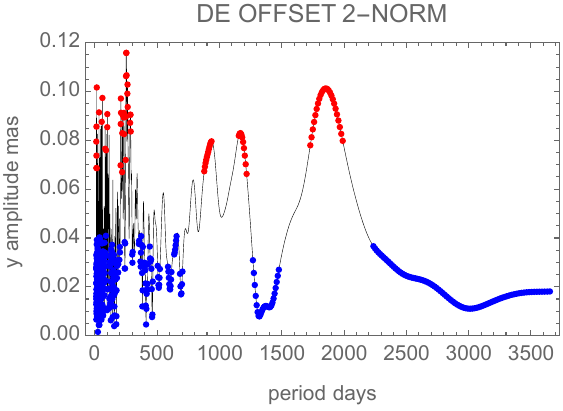}
\includegraphics[width=.45\textwidth]{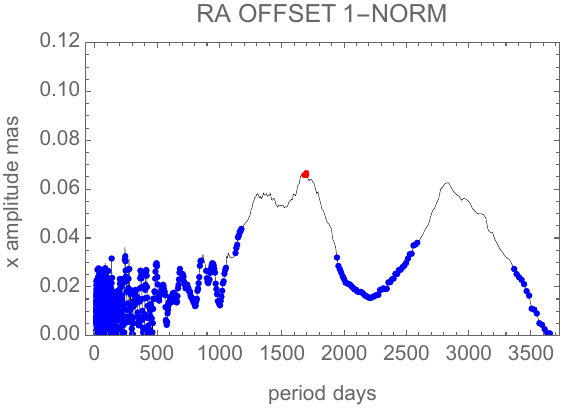}
\includegraphics[width=.45\textwidth]{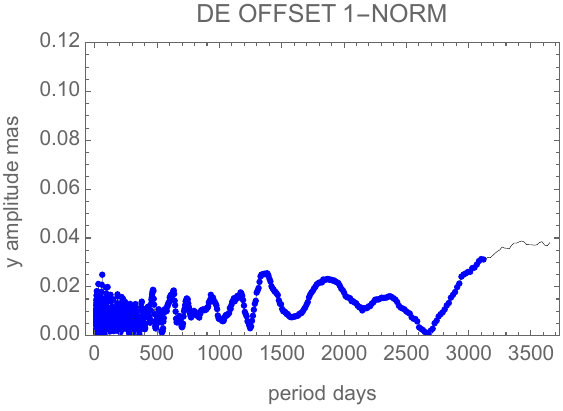}
\caption{Periodograms calculated for the astrometric time series shown in Fig. \ref{data.fig}. Left column: right ascension components in mas. Right column:
declination components in mas. Upper row: the classic (2-norm) unweighted LS periodogram. Lower row: the proposed robust 1-norm periodogram. 
In all graphs, the thin black curves represent computed periodogram amplitudes, the blue dots show the values below the 1-sigma confidence level,
the red dots show the values above the 3-sigma confidence level.\label{per.fig}}
\end{figure*}

The results are shown in the upper row plots of Fig. \ref{per.fig} for the two coordinate components. The periodogram amplitude estimates are connected with a black line to aid the eye. The significance of each periodogram point is also computed for this unweighted solution according to Sect. \ref{unc.sec}. We color-coded the significance by the normalized confidence level, so that estimates below the $1\sigma$ level are marked with blue dots, and estimates above the $3\sigma$ level are marked with red dots. A large number of values appear to be highly significant with periods across the entire range, including some short periods below 100 d, which obviously cannot be physical. This result, with a jungle of sharp peaks in the short-period domain and a few prominent features in the long-period domain, is typical of LS periodograms for ``noisy'' data. The inference is completely false, and we will now reveal why that happens.

\begin{figure*}
\centering
\includegraphics[width=.45\textwidth]{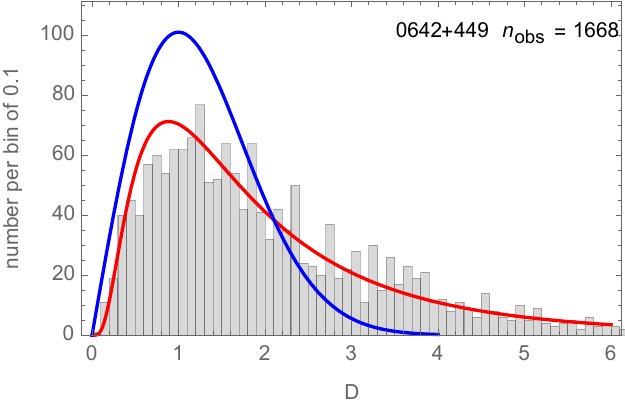}
\caption{Distribution of standardized astrometric deviations for the data set shown in Fig. \ref{data.fig}. The expected distribution, which is Rayleigh[1], is shown with the blue curve. The red curve is the empirical best-fitting distribution, which is LogNormal[0.484,0.779].}
\label{D.fig}
\end{figure*}

The single-epoch positions measured with VLBI are two-dimensional, and each position determination $\{x,y\}$ comes with a formal covariance $\boldsymbol{G}$, which is a 2 by 2 matrix. It is convenient to consider the normalized and centralized single-epoch position offset
\eb
D=\sqrt{(x-\bar x,y-\bar y)\,
\boldsymbol G^{-1}\,(x-\bar x,y-\bar y)^T},
\label{D.eq}
\ee
because it is a scalar variate, which is expected to follow a Rayleigh distribution with scale 1, reducing the dimensionality of statistical analysis to 1. The true coordinates $\{\bar x,\bar y\}$ are not known, but they can be separately estimated as the weighted mean position. Note that even the well-known formula for covariance $\boldsymbol{G}$ is based on the underlying assumption of normally distributed random errors. The variate $D$ allows us to test this basic assumption. The histogram of $D$ values computed for the example data set in Fig. \ref{data.fig} is shown in Fig. \ref{D.fig}. For reference, the expected Rayleigh[1] distribution (normalized to the same area) is shown with the blue line. We can see that the actual distribution of astrometric offsets is very far from the expectation, and the difference cannot be fixed just by scaling the formal errors. Although the mode of the empirical distribution is approximately where it is expected to be (at 1), a long and powerful tail stretching far beyond the Rayleigh[1] curve indicates that nearly half of the available measurements have values associated with nil probabilities of occurrence within the assumed statistical model.

The heavy-tailed nature of the data distribution invalidates the LS periodogram method. The data points with large deviations from the mean should not be called outliers in this case, because they represent a large part, if not the majority, of the population. Simple fixes such as clipping the data outside the $3\sigma$ threshold are not justified. The numerous deviant data points corrupt any least-squares estimation, and generate bogus signals in this periodogram analysis. Methods of robust estimation are designed to handle heavy-tailed data in a more consistent way. In particular, the 1-norm estimation seeks to minimize the sum of absolute values of residuals rather than the sum of their squares:
\eb 
\Lambda(\hat d)\equiv \sum_i |d_i-\hat d_i|={\rm min}.
\ee 
This merit function diminishes the impact of large deviants and permits a meaningful solution for any intrinsically symmetric populations. It is robust with respect to the subsample containing high normalized offsets, because each individual data point has an effectively lower weight in the solution, irrespective of its value. 

\section{Implementation of 1-norm periodograms}

The same periodogram models (Eqs. \ref{model.eq} and \ref{model2.eq}) can be used as in the classical LS method. The main differences in implementation are of the technical character. The main optimization problem is no longer linear, and it cannot be formalized as Eq. \ref{a.eq}. Consequently, there is no direct calculation of the associated covariance matrix of the periodogram coefficients. This can still be done numerically using Monte Carlo simulations. The solution itself is implemented with one of the existing global nonlinear optimization methods with vector-valued arguments, such as the Nelder-Mead (simplex downhill), differential evolution, or simulated annealing methods.\footnote{Cf. a useful summary in \url{https://reference.wolfram.com/language/tutorial/ConstrainedOptimizationGlobalNumerical.html}} These methods are computationally much more expensive than the regular LS periodogram. However, we achieved a computing time of about 1 min on a regular laptop for the given example with 1668 data points and 1000 trial periods for the two time series. 

The resulting 1-norm periodograms for the given data sets are shown in Fig. \ref{per.fig}, lower row. They are expressed in the same values (amplitudes, per Eq. \ref{amp.eq}) and units as the LS periodograms in the upper row, so that they can be directly compared. Formal confidence levels cannot be directly computed for the 1-norm solutions, because the population distribution is non-Gaussian. We reproduce, however, the re-normalized confidence intervals $1\sigma$ and $3\sigma$ from the LS solution to emphasize the significance of the results. 

Quite clearly, the robust 1-norm periodograms paint a different picture about the temporal variations of the given data. The amplitude values dropped by half or more, and most of the estimated values are now below the $1\sigma$-interval. The largest reduction is seen in the high-frequency domain. Given the nature of the object under investigation, the low-frequency features are of special interest. We find the main features at different locations than with the LS method, and for the RA component, they clearly dominate the spectral power distribution. Intriguingly, there is a compact location around 1730 d with periodogram amplitudes above $3\sigma$, which was completely insignificant in the LS solution. To test if this point is indeed associated with a high level of confidence, a much more extensive bootstrapping or Monte Carlo simulations are required.

\begin{figure*}
\centering
\includegraphics[width=.45\textwidth]{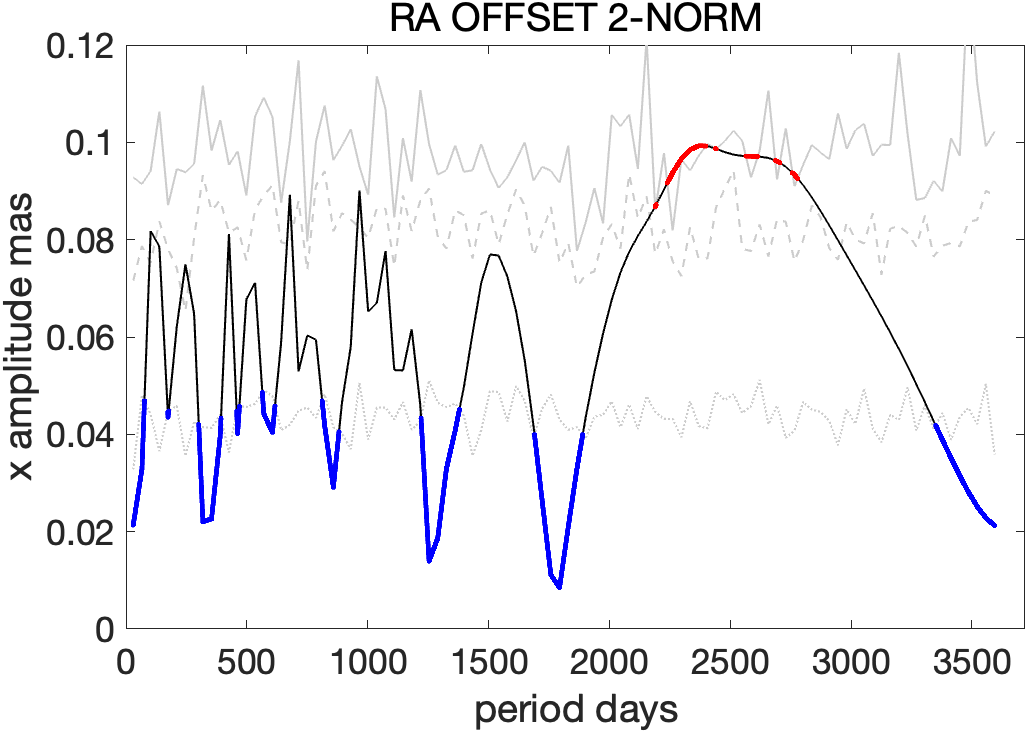}
\includegraphics[width=.45\textwidth]{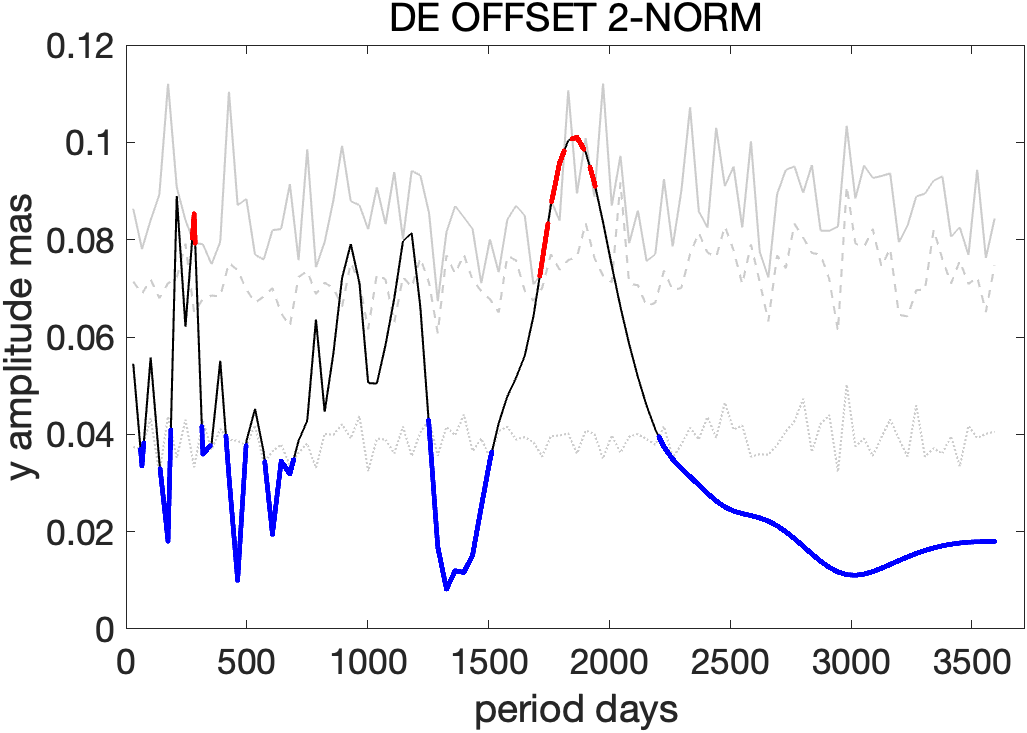}
\includegraphics[width=.45\textwidth]{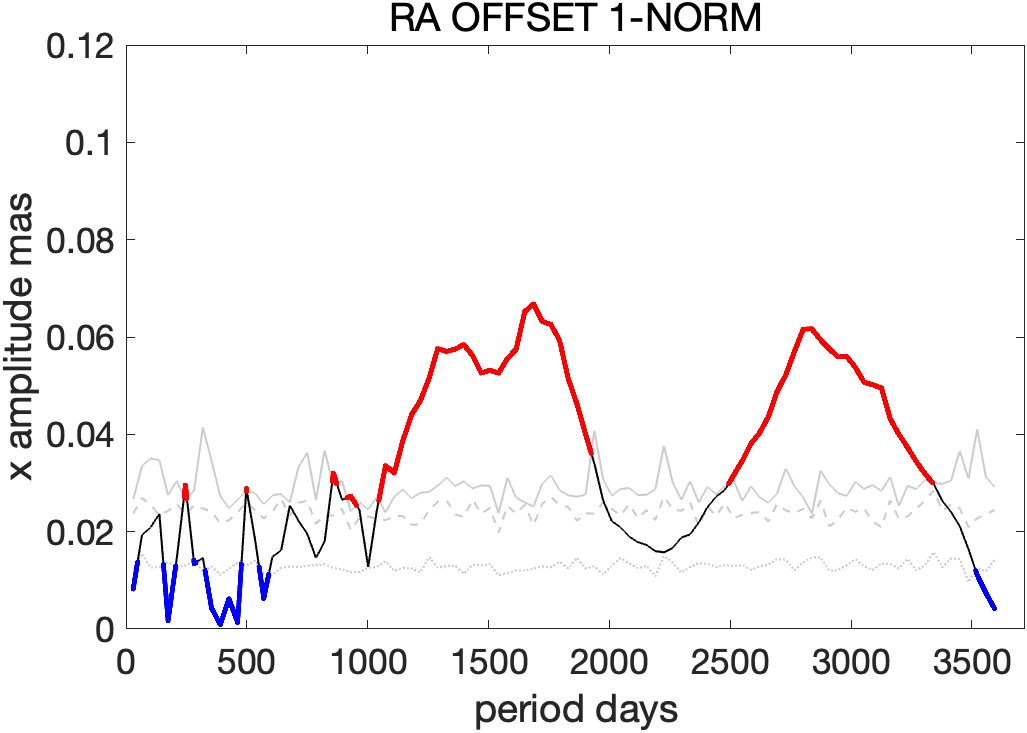}
\includegraphics[width=.45\textwidth]{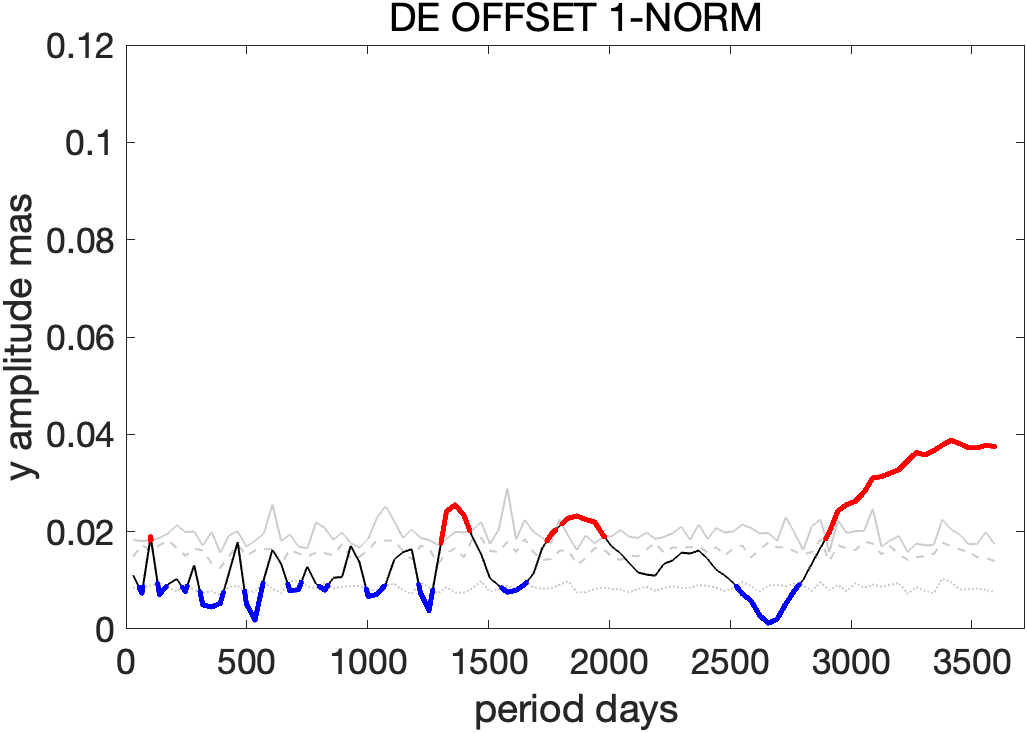}
\caption{Periodograms calculated for the astrometric time series shown in Fig. \ref{data.fig}. Left column: right ascension components in mas. Right column: declination components in mas. Upper row: the classic (2-norm) unweighted LS periodogram. Lower row: the proposed robust 1-norm periodogram. In all graphs, the thin black curves represent computed periodogram amplitudes, the blue dots show the values below the 68\% confidence level, the red dots show the values above the 99\% confidence level.\label{Boot.fig}}
\end{figure*}

We performed a non-parametric bootstrap by producing 100 data samples from the original time series by randomly permuting its elements but keeping the fixed cadence of epochs, thus inheriting the same distribution of the uncorrelated noise component as the original data. Then, for each data sample, we computed the LS and 1-norm periodograms. The bootstrap distribution appears reasonably symmetric. For each period, the $N$\% confidence interval ($0\le N\le100$) is given by the non-parametric percentile bootstrap interval contained between the $(N/2)$th and $(100-N/2)$th percentiles. For verification purposes, the periodogram analysis and the bootstrapping estimation was independently implemented by two authors using different computer languages, on the same data set. Figure~\ref{Boot.fig} displays the obtained results where the grey lines represents the difference between the upper and lower bounds of the confidence intervals, respectively for $N=68.3$\% (dotted line), $N=95.5$\% (dashed line), and $N=99.7$\% (solid line). In the case of the 2-norm implementation, the bootstrap provides a confidence interval consistent with results shown in Fig.~\ref{per.fig}. For the 1-norm spectrum, the confidence interval is lower than for the 2-norm, reflecting the robustness of the 1-norm estimation, which is less sensitive to the data points in the extended tail of the distribution. The amplitudes reach highest significant values, with a significance level well higher than 99.7\%, for the right ascension suggesting that the bump observed at a period of $\sim$1700~days is not due to chance and the secondary peak at $\sim$2800~days should also be considered for further investigations. We note, however, that the bootstrap-estimated confidence only refers to the random uncorrelated noise in the data (of arbitrary PDF). The data may include a time-correlated component of physical or instrumental origin. Broad periodogram features seen in Fig. \ref{Boot.fig} may require additional analysis using, e.g., structure functions of first or second order, which also incorporate periodic components with time-variable phase \citep{1978IEEEP..66.1048R, 1985ApJ...296...46S, 1991IEEEP..79..952R}.



\section{A search for periodic modulation in observed positions of ICRF3 sources}

Diurnal geodetic VLBI sessions have been regularly scheduled over nearly 40 years, using networks of stations separated by baselines of hundreds to thousands of kilometers long. Within each daily session, a number of widely separated radio sources are observed multiple times over the course of 24 hours. The resulting data are processed in a few data analysis centers, including the U.S. Naval Observatory. In this paper, we use a global solution for two-dimensional coordinates of epoch calculated at USNO (2022a) in the standard S/X band setup. This data product includes  astrometric time series from more than 6000 diurnal sessions. The total number of sources is 5153 in this data set, but here we only consider 259 of them with more than 200 single-epoch measurements.

\begin{deluxetable*}{r|RRr|c|c}
\tabletypesize{\scriptsize}
\tablewidth{0pt} 
\tablecaption{ICRF3 quasars with quasi-periodic signals \label{49.tab}}
\tablehead{
\colhead{IERS id} & \colhead{RA}& \colhead{Dec} & \colhead{$n_{\rm obs}$} & \colhead{Periods RA} & \colhead{Periods Dec}\\
\colhead{} & \colhead{$\degr$} & \colhead{$\degr$}& \colhead{} & \colhead{d} & \colhead{d}}
\startdata 
\text{0003-066} & 1.5578870361 & -6.3931487242 & 1648 & & 1760\\
 \text{0014+813} & 4.2853121110 & 81.5855934649 & 1272 & & $>3652$\\
 \text{0016+731} & 4.9407765018 & 73.4583382213 & 1214 & 1520, 1850, $>3652$ \\
 \text{0119+041} & 20.4869237314 & 4.3735373041 & 1717 & $>3652$ & \\
 \text{0119+115} & 20.4233126885 & 11.8306703578 & 1673 & & 1670, 2640 \\
 \text{0131-522} & 23.2740104059 & -52.0010959196 & 628 & 830 &\\
 \text{0202+149} & 31.2100578951 & 15.2364009985 & 963 & & 2940 \\
 \text{0300+470} & 45.8968425830 & 47.2711876296 & 825 & 1030 & multiple 946--3652\\
 \text{0308-611} & 47.4837464402 & -60.9775156729 & 1292 & & 800, 2070\\
 \text{0322+222} & 51.4033931446 & 22.4001015605 & 872 & 1660, $>3652$ & \\
 \text{0336-019} & 54.8789074465 & -1.7766122761 & 2403 & 1820 & \\
 \text{0537-286} & 84.9761728366 & -28.6655411473 & 556 & 2750 & \\
 \text{0537-441} & 84.7098398243 & -44.0858164005 & 2315 & 2760 & \\
 \text{0607-157} & 92.4206230931 & -15.7112979835 & 873 & 870, 1160 & \\
 \text{0637-752} & 98.9437829129 & -75.2713376525 & 607 & 990, 1110, 1550, 1970 & \\
 \text{0642+449} & 101.6334416364 & 44.8546083651 & 1668 & 1690 &\\
 \text{0735+178} & 114.5308072908 & 17.7052772667 & 502 & $>3652$ &\\
 \text{0748+126} & 117.7168572399 & 12.5180078358 & 942 & 1590, $>3652$ &\\
 \text{0749+540} & 118.2557690371 & 53.8832325243 & 1167 & $>3652$ & \\
 \text{0955+476} & 149.5819651824 & 47.4188451163 & 2879 & 1320 & \\
 \text{1030+415} & 158.2654494235 & 41.2683980499 & 310  & multiple 2230--3580 &\\
 \text{1039+811} & 161.0960939470 & 80.9109564077 & 239 & 1240, 2060, 2770 & \\
 \text{1123+264} & 171.4737996602 & 26.1722163092 & 345 & 3460 & \\
 \text{1213-172} & 183.9447990581 & -17.5292786545 & 373 & $>3652$ & \\
 \text{1308+326} & 197.6194327218 & 32.3454952446 & 2165 & & 2920 \\
 \text{1451-375} & 223.6142072756 & -37.7925402884 & 384 & 920, 1310, 1900, 2470 & 1030, 1310, 1900, 2530 \\
 \text{1546+027} & 237.3726535376 & 2.6169897834 & 619 & & 1980, 2180 \\
 \text{1548+056} & 237.6469551807 & 5.4529023152 & 313 & & 2380, 3180 \\
 \text{1611+343} & 243.4211010163 & 34.2133080213 & 2058 & & 3410 \\
 \text{1617+229} & 244.8117691700 & 22.7966252695 & 442 & & 3500 \\
 \text{1639-062} & 250.5090738210 & -6.3565819658 & 738 & 910 &\\
 \text{1639+230} & 250.3551148643 & 22.9511202107 & 619 & & 2360 \\
 \text{1739+522} & 265.1540743749 & 52.1953909459 & 2222 & 1350 & 2610 \\
 \text{1846+322} & 282.0920357219 & 32.3173899470 & 872 & 955, 1430, 3170 & 1370 \\
 \text{1849+670} & 282.3169678649 & 67.0949111894 & 469 & & 1390, 1680 \\
 \text{2007+777} & 301.3791605043 & 77.8786798684 & 263 & 1325 & \\
 \text{2128-123} & 322.8969239777 & -12.1179989354 & 890 & & 1890 \\
 \text{2136+141} & 324.7554552901 & 14.3933311595 & 1146 & 3240 & \\
 \text{2145+067} & 327.0227444590 & 6.9607233849 & 2182 & 3460 &\\
 \text{2201+315} & 330.8123991110 & 31.7606305396 & 948 & & 2015 \\
 \text{2214+350} & 334.0833745874 & 35.3039388710 & 750 & & 3430 \\
 \text{2216-038} & 334.7168238499 & -3.5935776334 & 435 & & 2420 \\
 \text{2223-052} & 336.4469137221 & -4.9503863240 & 1758 & 2290 & 1770 \\
 \text{2227-088} & 337.4170180620 & -8.5484543301 & 812 & 2690 & $>3652$ \\
 \text{2229+695} & 337.6519572538 & 69.7744658067 & 964 & 745, 1110, $>3652$ & \\
 \text{2232-488} & 338.8051524303 & -48.5996651465 & 209 & 540 & \\
 \text{2234+282} & 339.0936285247 & 28.4826147730 & 2265 & 1435, 2275 & 2275 \\
 \text{2243-123} & 341.5759665538 & -12.1142437955 & 1004 & & 1270, 1730 \\
 \text{2318+049} & 350.1869024698 & 5.2305423851 & 1230 & 2430 & 2430
\enddata
\tablecomments{IERS source names in column (1) preappend letter B to the codes as given. RA and Dec coordinates in cols. (2) and (3) are the weighted mean positions calculated from the given data, not the published ICRF3 coordinates. Periods in columns 5 and 6 are approximate local peaks in periodograms.}
\end{deluxetable*}

The 1-norm periodogram computation was uniformly applied to each of the frequently observed ICRF3 sources, separately for RA and Dec offsets from the weighted mean positions. These mean positions are specifically computed for the given data set and the solution version, so they may slightly differ from the published ICRF3 mean positions. The purpose of this numerical experiment was to identify sources with possible sinusoidal variations in the observed positions on the sky. The results are presented in a compact form in Table \ref{49.tab} for 49 quasars where the formal significance criterion $\psi>11.829$, from Eq. \ref{psi.eq}, is triggered in either of the coordinates for at least one trial period. The table provides IERS names of the sources (which should be predended with letter B to match Simbad identification), our computed mean RA and Dec coordinates in degrees, the total number of diurnal sessions, and the significant trial periods. 

The main result of this computation is that there seems to be no isolated single-frequency sinusoidal signals in the observational data similar to those that are found for astrometric binaries. Instead of well defined peaks in the periodograms, we find ``packages'' of trial periods with elevated amplitudes and significance levels. The emerging picture is more consistent with an ensemble of small vaguely periodic modes of non-commensurate frequencies. One-fifth of the detections have a rising periodogram amplitude toward the upper limit of this analysis (10 yr). The significant periods are mostly longer than 2 yr, although much shorter trial periods have been tested. Only 19\% of the sample show periodic variations above the $3\sigma$ level. The typical peak amplitudes are in the range $100$--$200$ $\mu$as.

We note that the robust periodograms were computed separately for the RA and Dec coordinates of epoch positions. In the case of Keplerian motion in a binary system, the detectable signal may be present in both coordinates with the same principal period (and its harmonics) but with different phase and amplitude. The probability density of the angle $i$ between the line of sight and the vector of orbital angular momentum is proportional to $|\sin(i)|$. Therefore, nearly face-on projected orbits are less likely than nearly edge-on orbits. For marginally detectable trajectories, the detactable signal is mostly present in one dimension, which is uniformly distributed with respect to the local north direction. The largest extent of the projected orbit can be aligned with one of the coordinate axes with the same probability as a tilt of $45\degr$ or $135\degr$. In the latter case, the detectable signal is split between the coordinates, and it should be harder to find it with confidence. Possible ways to deal with this problem include rotating the RA-Dec measurements on a grid of position angles to find a preferred direction maximizing the signal amplitude from a 1D periodogram. Technically, if a significant single-period signal is detected in both RA and Dec coordinates, a 2D version of robust periodogram can be implemented. The available coordinate measurements are combined in a single LS adjustment, but the number of unknown terms per trial period increases to a minimum of 8 because of the unknown phase. This may erode the confidence level of the signal, if the the projected orbit is strongly elongated due to the geometric orientation or large eccentricity. Quasars IERS B1451$-$375, 2234$+$282, and 2318$+049$ are attractive targets for further investigation, because they show coherent periodicities in both coordinates from our results in Table \ref{49.tab}.

\section{Summary and discussion}
\label{sum.sec}

We have shown in this paper that the classical LS periodogram method is firmly based on strong and restrictive assumptions about the distribution of post-fit residuals (which is assumed to be Normal) and the character of physical signals in the data. It provides an optimal, unbiased, and unique solution for periodogram power or amplitude only under these conditions. Whenever the sample distribution shows significant departures from the Gaussian PDF, or more complex signals are present that are not captured in the model, the LS method becomes corrupted and can produce absolutely misleading results.

We have considered a specific observational data set for a moderately variable ICRF3 source collected over $>30$ yr by the global geodetic VLBI system. The distribution of astrometric positional offsets with respect to the mean position on the sky is explicitly non-Gaussian when scaled with the given 2D formal covariances or in absolute values. The normalized offsets are well represented by a log-normal distribution with a tighter mode and a heavy tail extending to high values. Nearly half of the measurements are way outside of the expected distribution. As a result, the traditional 2-norm (LS) periodogram produces a complex structure with multiple features that are formally above the $3\sigma$ confidence interval across the spectrum of trial periods. This result is completely bogus. The robust 1-norm periodogram method, when applied to the same data, produces amplitudes that are smaller by half or more. Ranked by the same previously estimated single-point confidence, the 1-norm values are all insignificant except for a single point in the RA component with a period of 1730 d and amplitude 72 $\mu$as, which appears to be above $3\sigma$. Is this periodic signal real? The best way to find out is to continue taking high-precision measurements of this source with VLBI for a few years. A stable sinusoidal signal, which could be produced by an orbiting binary black hole, for example, would emerge more strongly on the longer time scale. Alternatively, physical models could be tested, where transient periodic signals wax and wane in segments of the data due to phase scrambling. This new method provides the opportunity to more reliably and extensively search for periodic signals in non-Gaussian time series at the margin of available accuracy.

The 1-norm periodogram computation was performed for 259 ICRF3 sources with more than 200 diurnal sessions collected over nearly 40 years. These measurements are characterized by heavy-tailed sample distributions of residuals. We identified 49 objects (19\%), which have at least one statistically significant periodogram value in either coordinate component. Short periods are never found, indicating a possible physical mechanism of these signals in the transient structure of the radio-emitting sources. The signals are not consistent with clean sinusoidal variation at a specific frequency, which would emerge for an orbiting binary black hole. Rather, the pattern is that of ``vague periodicity'' represented by packages of sine waves with a distribution of frequencies. A possible physical model is a source that moves in loops on the sky returning to the vicinity of the initial position after some characteristic time, which may also vary with time. The estimated amplitude of these vaguely periodic excursions is 70 $\mu$as and higher. About one-fifth of the detected signals are truncated by the upper boundary of our periodograms (10 yr). Further investigation of these astrometric wobbles and continuous daily measurements will refine the models and allow us to understand the nature of the phenomenon.

\section*{Supplementary Materials}
A python code and a demonstration notebook implementing the robust L1 periodogram algorithm and confidence (or false alarm probability) computation with sample random permutation was published by A. Goldin in \url{https://github.com/agoldin/LombScargleL1} and \url{https://github.com/agoldin/LombScargleL1/blob/main/demo_comparison.ipynb}.


\vspace{5mm}
\facilities{VLBI}


\software{  
          Wolfram Mathematica
          }

\bibliography{main}

@ARTICLE{2015ApJ...813L..41A,
       author = {{Ackermann}, M. and {Ajello}, M. and {Albert}, A. and {Atwood}, W.~B. and {Baldini}, L. and {Ballet}, J. and {Barbiellini}, G. and {Bastieri}, D. and {Becerra Gonzalez}, J. and {Bellazzini}, R. and {Bissaldi}, E. and {Blandford}, R.~D. and {Bloom}, E.~D. and {Bonino}, R. and {Bottacini}, E. and {Bregeon}, J. and {Bruel}, P. and {Buehler}, R. and {Buson}, S. and {Caliandro}, G.~A. and {Cameron}, R.~A. and {Caputo}, R. and {Caragiulo}, M. and {Caraveo}, P.~A. and {Cavazzuti}, E. and {Cecchi}, C. and {Chekhtman}, A. and {Chiang}, J. and {Chiaro}, G. and {Ciprini}, S. and {Cohen-Tanugi}, J. and {Conrad}, J. and {Cutini}, S. and {D'Ammando}, F. and {de Angelis}, A. and {de Palma}, F. and {Desiante}, R. and {Di Venere}, L. and {Dom{\'\i}nguez}, A. and {Drell}, P.~S. and {Favuzzi}, C. and {Fegan}, S.~J. and {Ferrara}, E.~C. and {Focke}, W.~B. and {Fuhrmann}, L. and {Fukazawa}, Y. and {Fusco}, P. and {Gargano}, F. and {Gasparrini}, D. and {Giglietto}, N. and {Giommi}, P. and {Giordano}, F. and {Giroletti}, M. and {Godfrey}, G. and {Green}, D. and {Grenier}, I.~A. and {Grove}, J.~E. and {Guiriec}, S. and {Harding}, A.~K. and {Hays}, E. and {Hewitt}, J.~W. and {Hill}, A.~B. and {Horan}, D. and {Jogler}, T. and {J{\'o}hannesson}, G. and {Johnson}, A.~S. and {Kamae}, T. and {Kuss}, M. and {Larsson}, S. and {Latronico}, L. and {Li}, J. and {Li}, L. and {Longo}, F. and {Loparco}, F. and {Lott}, B. and {Lovellette}, M.~N. and {Lubrano}, P. and {Magill}, J. and {Maldera}, S. and {Manfreda}, A. and {Max-Moerbeck}, W. and {Mayer}, M. and {Mazziotta}, M.~N. and {McEnery}, J.~E. and {Michelson}, P.~F. and {Mizuno}, T. and {Monzani}, M.~E. and {Morselli}, A. and {Moskalenko}, I.~V. and {Murgia}, S. and {Nuss}, E. and {Ohno}, M. and {Ohsugi}, T. and {Ojha}, R. and {Omodei}, N. and {Orlando}, E. and {Ormes}, J.~F. and {Paneque}, D. and {Pearson}, T.~J. and {Perkins}, J.~S. and {Perri}, M. and {Pesce-Rollins}, M. and {Petrosian}, V. and {Piron}, F. and {Pivato}, G. and {Porter}, T.~A. and {Rain{\`o}}, S. and {Rando}, R. and {Razzano}, M. and {Readhead}, A. and {Reimer}, A. and {Reimer}, O. and {Schulz}, A. and {Sgr{\`o}}, C. and {Siskind}, E.~J. and {Spada}, F. and {Spandre}, G. and {Spinelli}, P. and {Suson}, D.~J. and {Takahashi}, H. and {Thayer}, J.~B. and {Thompson}, D.~J. and {Tibaldo}, L. and {Torres}, D.~F. and {Tosti}, G. and {Troja}, E. and {Uchiyama}, Y. and {Vianello}, G. and {Wood}, K.~S. and {Wood}, M. and {Zimmer}, S. and {Berdyugin}, A. and {Corbet}, R.~H.~D. and {Hovatta}, T. and {Lindfors}, E. and {Nilsson}, K. and {Reinthal}, R. and {Sillanp{\"a}{\"a}}, A. and {Stamerra}, A. and {Takalo}, L.~O. and {Valtonen}, M.~J.},
        title = "{Multiwavelength Evidence for Quasi-periodic Modulation in the Gamma-Ray Blazar PG 1553+113}",
      journal = {\apjl},
         year = 2015,
        month = nov,
       volume = {813},
       number = {2},
          eid = {L41},
        pages = {L41},
          doi = {10.1088/2041-8205/813/2/L41},
archivePrefix = {arXiv},
       eprint = {1509.02063},
 primaryClass = {astro-ph.HE},
       adsurl = {https://ui.adsabs.harvard.edu/abs/2015ApJ...813L..41A}
}

@ARTICLE{2008MNRAS.385.1279B,
       author = {{Baluev}, R.~V.},
        title = "{Assessing the statistical significance of periodogram peaks}",
      journal = {\mnras},
         year = 2008,
        month = apr,
       volume = {385},
       number = {3},
        pages = {1279-1285},
          doi = {10.1111/j.1365-2966.2008.12689.x},
archivePrefix = {arXiv},
       eprint = {0711.0330},
 primaryClass = {astro-ph},
       adsurl = {https://ui.adsabs.harvard.edu/abs/2008MNRAS.385.1279B}
}

@ARTICLE{2020A&A...644A.159C,
       author = {{Charlot}, P. and {Jacobs}, C.~S. and {Gordon}, D. and {Lambert}, S. and {de Witt}, A. and {B{\"o}hm}, J. and {Fey}, A.~L. and {Heinkelmann}, R. and {Skurikhina}, E. and {Titov}, O. and {Arias}, E.~F. and {Bolotin}, S. and {Bourda}, G. and {Ma}, C. and {Malkin}, Z. and {Nothnagel}, A. and {Mayer}, D. and {MacMillan}, D.~S. and {Nilsson}, T. and {Gaume}, R.},
        title = "{The third realization of the International Celestial Reference Frame by very long baseline interferometry}",
      journal = {\aap},
         year = 2020,
        month = dec,
       volume = {644},
          eid = {A159},
        pages = {A159},
          doi = {10.1051/0004-6361/202038368},
archivePrefix = {arXiv},
       eprint = {2010.13625},
 primaryClass = {astro-ph.GA},
       adsurl = {https://ui.adsabs.harvard.edu/abs/2020A&A...644A.159C}
}

@ARTICLE{1999ApJ...526..890C,
       author = {{Cumming}, Andrew and {Marcy}, Geoffrey W. and {Butler}, R. Paul},
        title = "{The Lick Planet Search: Detectability and Mass Thresholds}",
      journal = {\apj},
         year = 1999,
        month = dec,
       volume = {526},
       number = {2},
        pages = {890-915},
          doi = {10.1086/308020},
archivePrefix = {arXiv},
       eprint = {astro-ph/9906466},
 primaryClass = {astro-ph},
       adsurl = {https://ui.adsabs.harvard.edu/abs/1999ApJ...526..890C}
}

@ARTICLE{1981AJ.....86..619F,
       author = {{Ferraz-Mello}, S.},
        title = "{Estimation of Periods from Unequally Spaced Observations}",
      journal = {\aj},
         year = 1981,
        month = apr,
       volume = {86},
        pages = {619},
          doi = {10.1086/112924},
       adsurl = {https://ui.adsabs.harvard.edu/abs/1981AJ.....86..619F}
}

@ARTICLE{2023AJ....165..202F,
       author = {{Frouard}, Julien},
        title = "{Robust Estimates of Orientation between Astrometric Catalogs}",
      journal = {\aj},
         year = 2023,
        month = may,
       volume = {165},
       number = {5},
          eid = {202},
        pages = {202},
          doi = {10.3847/1538-3881/acc6cb},
       adsurl = {https://ui.adsabs.harvard.edu/abs/2023AJ....165..202F}
}

@ARTICLE{2006ApJS..166..341G,
       author = {{Goldin}, A. and {Makarov}, V.~V.},
        title = "{Unconstrained Astrometric Orbits for Hipparcos Stars with Stochastic Solutions}",
      journal = {\apjs},
         year = 2006,
        month = sep,
       volume = {166},
       number = {1},
        pages = {341-350},
          doi = {10.1086/505939},
archivePrefix = {arXiv},
       eprint = {astro-ph/0606293},
 primaryClass = {astro-ph},
       adsurl = {https://ui.adsabs.harvard.edu/abs/2006ApJS..166..341G}
}

@ARTICLE{2023AnRSA..10..623H,
       author = {{Hara}, Nathan C. and {Ford}, Eric B.},
        title = "{Statistical Methods for Exoplanet Detection with Radial Velocities}",
      journal = {Annual Review of Statistics and Its Application},
         year = 2023,
        month = mar,
       volume = {10},
       number = {1},
        pages = {623-649},
          doi = {10.1146/annurev-statistics-033021-012225},
archivePrefix = {arXiv},
       eprint = {2308.00701},
 primaryClass = {astro-ph.IM},
       adsurl = {https://ui.adsabs.harvard.edu/abs/2023AnRSA..10..623H}
}

@ARTICLE{2018ApJS..236...16V,
       author = {{VanderPlas}, Jacob T.},
        title = "{Understanding the Lomb-Scargle Periodogram}",
      journal = {\apjs},
         year = 2018,
        month = may,
       volume = {236},
       number = {1},
          eid = {16},
        pages = {16},
          doi = {10.3847/1538-4365/aab766},
archivePrefix = {arXiv},
       eprint = {1703.09824},
 primaryClass = {astro-ph.IM},
       adsurl = {https://ui.adsabs.harvard.edu/abs/2018ApJS..236...16V}
}

@ARTICLE{2023A&A...669A.138L,
       author = {{Lambert}, S. and {Malkin}, Z.},
        title = "{Estimation of large-scale deformations in VLBI radio source catalogs with mitigation of impact of outliers: A comparison between different L1- and L2-norm-based methods}",
      journal = {\aap},
         year = 2023,
        month = jan,
       volume = {669},
          eid = {A138},
        pages = {A138},
          doi = {10.1051/0004-6361/202244837},
       adsurl = {https://ui.adsabs.harvard.edu/abs/2023A&A...669A.138L}
}

@ARTICLE{1976Ap&SS..39..447L,
       author = {{Lomb}, N.~R.},
        title = "{Least-Squares Frequency Analysis of Unequally Spaced Data}",
      journal = {\apss},
         year = 1976,
        month = feb,
       volume = {39},
       number = {2},
        pages = {447-462},
          doi = {10.1007/BF00648343},
       adsurl = {https://ui.adsabs.harvard.edu/abs/1976Ap&SS..39..447L}
}

@ARTICLE{2009ApJ...707L..73M,
       author = {{Makarov}, V.~V. and {Beichman}, C.~A. and {Catanzarite}, J.~H. and {Fischer}, D.~A. and {Lebreton}, J. and {Malbet}, F. and {Shao}, M.},
        title = "{Starspot Jitter in Photometry, Astrometry, and Radial Velocity Measurements}",
      journal = {\apjl},
         year = 2009,
        month = dec,
       volume = {707},
       number = {1},
        pages = {L73-L76},
          doi = {10.1088/0004-637X/707/1/L73},
archivePrefix = {arXiv},
       eprint = {0911.2008},
 primaryClass = {astro-ph.SR},
       adsurl = {https://ui.adsabs.harvard.edu/abs/2009ApJ...707L..73M}
}

@ARTICLE{2010ApJ...717.1202M,
       author = {{Makarov}, V.~V. and {Parker}, D. and {Ulrich}, R.~K.},
        title = "{Astrometric Jitter of the Sun as a Star}",
      journal = {\apj},
         year = 2010,
        month = jul,
       volume = {717},
       number = {2},
        pages = {1202-1205},
          doi = {10.1088/0004-637X/717/2/1202},
archivePrefix = {arXiv},
       eprint = {1005.4888},
 primaryClass = {astro-ph.SR},
       adsurl = {https://ui.adsabs.harvard.edu/abs/2010ApJ...717.1202M}
}

@ARTICLE{2012MmSAI..83..952M,
       author = {{Makarov}, V. and {Berghea}, C. and {Boboltz}, D. and {Dieck}, C. and {Dorland}, B. and {Dudik}, R. and {Fey}, A. and {Gaume}, R. and {Lei}, X. and {Schmitt}, H. and {Zacharias}, N.},
        title = "{Quasometry, its use and purpose}",
      journal = {\memsai},
         year = 2012,
        month = jan,
       volume = {83},
        pages = {952},
          doi = {10.48550/arXiv.1202.5283},
archivePrefix = {arXiv},
       eprint = {1202.5283},
 primaryClass = {astro-ph.IM},
       adsurl = {https://ui.adsabs.harvard.edu/abs/2012MmSAI..83..952M}
}

@ARTICLE{2017ApJ...845..149M,
       author = {{Makarov}, Valeri V. and {Goldin}, Alexey},
        title = "{Kepler Data on KIC 7341653: A Nearby M Dwarf with Monster Flares and a Phase-coherent Variability}",
      journal = {\apj},
         year = 2017,
        month = aug,
       volume = {845},
       number = {2},
          eid = {149},
        pages = {149},
          doi = {10.3847/1538-4357/aa7d06},
archivePrefix = {arXiv},
       eprint = {1707.00538},
 primaryClass = {astro-ph.SR},
       adsurl = {https://ui.adsabs.harvard.edu/abs/2017ApJ...845..149M}
}

@ARTICLE{2021MNRAS.506.5540M,
       author = {{Malkin}, Zinovy},
        title = "{Towards a robust estimation of orientation parameters between ICRF and Gaia celestial reference frames}",
      journal = {\mnras},
         year = 2021,
        month = oct,
       volume = {506},
       number = {4},
        pages = {5540-5547},
          doi = {10.1093/mnras/stab2100},
archivePrefix = {arXiv},
       eprint = {2107.08967},
 primaryClass = {astro-ph.IM},
       adsurl = {https://ui.adsabs.harvard.edu/abs/2021MNRAS.506.5540M}
}

@ARTICLE{1978IEEEP..66.1048R,
       author = {{Rutman}, J.},
        title = "{Characterization of phase and frequency instabilities in precision frequency sources: fifteen years of progress.}",
      journal = {IEEE Proceedings},
         year = 1978,
        month = sep,
       volume = {66},
        pages = {1048-1075},
       adsurl = {https://ui.adsabs.harvard.edu/abs/1978IEEEP..66.1048R}
}

@ARTICLE{1991IEEEP..79..952R,
       author = {{Rutman}, Jacques and {Walls}, F.~L.},
        title = "{Characterization of frequency stability in precision frequency sources.}",
      journal = {IEEE Proceedings},
         year = 1991,
        month = jul,
       volume = {79},
        pages = {952-960},
       adsurl = {https://ui.adsabs.harvard.edu/abs/1991IEEEP..79..952R}
}

@ARTICLE{1982ApJ...263..835S,
       author = {{Scargle}, J.~D.},
        title = "{Studies in astronomical time series analysis. II. Statistical aspects of spectral analysis of unevenly spaced data.}",
      journal = {\apj},
         year = 1982,
        month = dec,
       volume = {263},
        pages = {835-853},
          doi = {10.1086/160554},
       adsurl = {https://ui.adsabs.harvard.edu/abs/1982ApJ...263..835S}
}

@ARTICLE{1985ApJ...296...46S,
       author = {{Simonetti}, J.~H. and {Cordes}, J.~M. and {Heeschen}, D.~S.},
        title = "{Flicker of extragalactic radio sources at two frequencies.}",
      journal = {\apj},
         year = 1985,
        month = sep,
       volume = {296},
        pages = {46-59},
          doi = {10.1086/163418},
       adsurl = {https://ui.adsabs.harvard.edu/abs/1985ApJ...296...46S}
}

@ARTICLE{2014MNRAS.440.2099S,
       author = {{S{\"u}veges}, M.},
        title = "{Extreme-value modelling for the significance assessment of periodogram peaks}",
      journal = {\mnras},
         year = 2014,
        month = may,
       volume = {440},
       number = {3},
        pages = {2099-2114},
          doi = {10.1093/mnras/stu372},
       adsurl = {https://ui.adsabs.harvard.edu/abs/2014MNRAS.440.2099S}
}

@ARTICLE{2016AJ....152..151X,
       author = {{Xu}, Ming H. and {Heinkelmann}, Robert and {Anderson}, James M. and {Mora-Diaz}, Julian and {Schuh}, Harald and {Wang}, Guang L.},
        title = "{The Source Structure of 0642+449 Detected from the CONT14 Observations}",
      journal = {\aj},
         year = 2016,
        month = nov,
       volume = {152},
       number = {5},
          eid = {151},
        pages = {151},
          doi = {10.3847/0004-6256/152/5/151},
archivePrefix = {arXiv},
       eprint = {1607.04706},
 primaryClass = {astro-ph.IM},
       adsurl = {https://ui.adsabs.harvard.edu/abs/2016AJ....152..151X}
}
\bibliographystyle{aasjournal}

\end{document}